\documentclass[12pt,reqno]{article}
\usepackage{amsmath}

\allowdisplaybreaks
\numberwithin{equation}{section}

\newcommand{\I}{\mathrm{i}}
\newcommand{\e}{\mathrm{e}}

\newcommand{\ep}{\varepsilon}
\newcommand{\si}{\sigma}
\newcommand{\p}{\partial}
\newcommand{\half}{\frac{1}{2}}

\DeclareSymbolFont{AMSa}{U}{msa}{m}{n}
\DeclareSymbolFont{AMSb}{U}{msb}{m}{n}
\DeclareMathSymbol{\fieldR}{\mathalpha}{AMSb}{"52}

\begin{document} 

\begin{flushright} \small
 ITP--UU--03/45 \\ SPIN--03/28 \\ TUW--03--27 \\ hep-th/0309220
\end{flushright}
\bigskip

\begin{center}
 {\large\bfseries Instanton Solutions for the Universal 
Hypermultiplet}\footnote[2]{Talk given by S.~Vandoren. 
To appear in the proceedings of the 36th International Symposium
Ahrenshoop on the Theory of Elementary Particles, August 2003.} \\[5mm]
Marijn Davidse$^1$, Mathijs de Vroome$^1$,
Ulrich Theis$^2$ and Stefan Vandoren$^1$ \\[3mm]
 {\small\slshape
 $^1$Institute for Theoretical Physics \emph{and} Spinoza Institute \\
 Utrecht University, 3508 TD Utrecht, The Netherlands \\
 {\upshape\ttfamily M.Davidse, M.T.deVroome, S.Vandoren@phys.uu.nl}\\[3mm]
 $^2$Institute for Theoretical Physics, Vienna University of
 Technology, \\ Wiedner Hauptstrasse 8--10, A-1040 Vienna, Austria \\
 {\upshape\ttfamily theis@hep.itp.tuwien.ac.at}}
\end{center}
\vspace{5mm}

\hrule\bigskip

\centerline{\bfseries Abstract} \medskip

We expand our previous analysis on fivebrane and membrane instanton
solutions in the universal hypermultiplet, including near-extremal
multi-centered solutions and mixed five\-brane-membrane charged
instantons. The results are most conveniently described in terms of a
double-tensor multiplet.
\bigskip

\hrule\bigskip

\section{Introduction}

Low energy effective actions for type II strings on a Calabi-Yau (CY)
threefold are determined by $D=4$ $N=2$ supergravity actions coupled to
vector- and hypermultiplets, or multiplets that can be dualized into
these. The number of such multiplets depends on the Hodge numbers of the
CY, and is respectively $h_{1,1}$ $(h_{1,2})$ and $h_{1,2}+1$
$(h_{1,1}+1)$ for type IIA(B). The hypermultiplet moduli space is
quaternion-K\"ahler and contains the dilaton. Therefore this space
receives quantum corrections, both perturbatively and non-perturbatively.
In this note, we focus on rigid ($h_{1,2}=0$) CY threefold
compactifications of type IIA, or its mirror version in type IIB,
understood in terms of a Landau-Ginzburg orbifold \cite{AG}. Then there
is only a single hypermultiplet, the universal hypermultiplet, whose
tree level effective action is determined by the quaternion-K\"ahler
target space \cite{CFG,FS}
 \begin{equation} \label{UHM-QK}
  {\cal M}_H = \frac{\mathrm{SU}(1,2)}{\mathrm{U}(2)}\ .
 \end{equation}
This particular coset space is also K\"ahler, and there exist complex 
coordinates $S$ and $C$ in which the K\"ahler potential is given by
\begin{equation}
K=-{\rm ln} (S+\bar S -2C\bar C)\ .
\end{equation}
An alternative parametrization of the universal hypermultiplet in terms
of four real scalars is
 \begin{equation}
  e^{-1} \mathcal{L}_\mathrm{UH} = 
	- \frac{1}{2}\, \p^\mu \phi\, \p_\mu \phi - \frac{1}{2}\,
	\e^{-\phi} \big( \p^\mu \chi\, \p_\mu \chi + \p^\mu \varphi\,
	\p_\mu \varphi \big) 
  - \frac{1}{2}\, \e^{-2\phi} \big( \p_\mu \si + \chi \p_\mu
	\varphi \big)^2 \ .
 \end{equation}
The relation with the complex coordinates above is given by
\begin{eqnarray}\label{real-var}
 \e^\phi=\half (S+\bar S - 2 C \bar{C}) & \qquad & \chi = C+\bar C\
 ,\nonumber\\
 \sigma=\frac{\I}{2}(S-\bar S + C^2 - \bar{C}^2) & \qquad & \varphi =
 -\I (C-\bar C)\ .
\end{eqnarray}
Recently, the perturbative analysis of \cite{S,GHL} was revisited in 
\cite{AMTV}, and it was shown that non-trivial perturbative quantum
corrections only appear at one-loop. We will not discuss these
corrections here, but only mention that they were found by studying 
deformations of (\ref{UHM-QK}) that preserve the Heisenberg
subgroup of the classical group of SU(1,2) isometries
\begin{equation}
S\rightarrow S +\I\alpha + 2\bar\epsilon\, C + |\epsilon|^2\ ,\qquad
C\rightarrow C+\epsilon\ .
\end{equation}
The parameters $\alpha$ and $\epsilon$ are real and complex respectively
and, under the assumption that these transformations do not receive any
quantum corrections, generate the symmetries that are preserved in
string perturbation theory \cite{S}. In the basis of the real variables
(\ref{real-var}), the Heisenberg algebra of infinitesimal
transformations is generated by
\begin{equation}\label{H-A}
\delta \phi =0\ ,\qquad  \delta \chi = \epsilon + \bar \epsilon\ ,\qquad
\delta \varphi = -\I (\epsilon -\bar \epsilon)\ ,\qquad 
\delta \sigma = -\alpha -  (\epsilon + \bar \epsilon )\varphi\ .
\end{equation}
One can therefore choose a set of two commuting isometries,
corresponding to the parameters $\alpha$ and $\epsilon=\I\beta$ (with
$\beta$ real), to dualize the pseudoscalars $\varphi$ and $\sigma$ into
two tensors, using the known Legendre transformation techniques. In such
a double-tensor multiplet formulation, perturbative corrections are
highly constrained due to the rather restrictive couplings of
scalar-tensor systems with $N=2$ supersymmetry \cite{TV2}.

After dualization, the resulting tree level double-tensor multiplet 
Lagrangian reads \cite{TV1,TV2}
 \begin{equation} \label{DTM-action}
  e^{-1}\mathcal{L}_\mathrm{DT} = - \half\, \p^\mu \phi\, \p_\mu \phi
  - \half\,  \e^{-\phi}\, \p^\mu \chi\, \p_\mu \chi + \half M^{IJ}
  H^\mu_I  H_{\mu J}\ ,
 \end{equation}
where the $H_I$ are a pair of three-form field strengths, $H^\mu_I=\half
\varepsilon^{\mu\nu\rho\sigma}\partial_\nu B_{\rho\sigma\,I}$, and
 \begin{equation}
  M = \e^{\phi} \begin{pmatrix} 1 & - \chi \cr - \chi & \e^{\phi}
  + \chi^2 \end{pmatrix}\ .
 \end{equation}
The two scalars $\phi$ and $\chi$ parameterize the coset SL$(2,\fieldR)/
\mathrm{O}(2)$. The presence of the tensors breaks the SL$(2,\fieldR)$
symmetries to a two-dimensional subgroup generated by a certain
rescaling of the fields and by the remaining generator of the Heisenberg
algebra (\ref{H-A}) with real parameter $\epsilon=\gamma/2$. It acts as
a shift on $\chi$ and transforms the tensors linearly into each other
\cite{TV1},
 \begin{equation}
  \chi \rightarrow \chi + \gamma\ ,\qquad B_1 \rightarrow B_1 + \gamma
  B_2\ ,
 \end{equation}
with $\phi$ and $B_2$ invariant. An invariant combination is then the
tensor $\hat{H}_1=H_1-\chi H_2$, and we will call this the Heisenberg
invariant. An off-shell superspace formulation of the Lagrangian
(\ref{DTM-action}) was given in \cite{DWRV}, in terms of a single
function satisfying a linear second order differential equation. In that
paper, the superconformal calculus was used to write down the action.
Upon gauge-fixing the conformal symmetries, it yields (\ref{DTM-action})
or, equivalently, the action given in terms of the Calderbank-Pedersen
variables \cite{CP}.

In this note, we elaborate on \cite{TV1}, and investigate the
non-perturbative effects that contribute to the low energy effective
action. As explained in \cite{BBS}, these arise from Euclidean branes
wrapped around supersymmetric cycles in the CY\@. In the case of IIA
with $h_{1,2}=0$, the NS-fivebrane can wrap the entire CY, or the
D2-brane can wrap a non-trivial three-cycle inside the CY\@. From the
four-dimensional point of view, such configurations are localized in
space and time and correspond to fivebrane and membrane instantons of
the $N=2$ $D=4$ supergravity action.

In \cite{TV1}, following previous work of \cite{GS}, the Bogomol'nyi
equations were derived from (\ref{DTM-action}). The solutions were shown
to describe fivebrane and membrane-like instantons, and we review and
extend this below. The natural description of these instantons was given
in terms of the Euclidean continuation of the double-tensor multiplet
action. This has the advantage over the Euclidean universal
hypermultiplet that the Euclidean action is semi-positive definite and
hence it justifies the semi-classical approximation \cite{TV1}.
Perturbatively, the double-tensor multiplet guarantees
U$(1)\times\mathrm{U}(1)$ isometries in the dual hypermultiplet
description. Non-perturbatively, however, the duality is expected to
involve also the constant modes of the dual scalars $\varphi$ and
$\sigma$ by means of theta-angle-like terms. Such terms break the U(1)
isometries to a discrete subgroup, see e.g. \cite{OV,K}.

\section{Fivebrane Instantons}

The first Bogomol'nyi equation that can be derived from the
double-tensor multiplet action is given by \cite{TV1}
 \begin{equation} \label{5-inst}
  \begin{pmatrix} H_{\mu1} \cr H_{\mu2}\end{pmatrix} = \pm\, \p_\mu 
\begin{pmatrix} \e^{-\phi}
  \chi \cr \e^{-\phi}\end{pmatrix}\ ,
 \end{equation}
where the plus and minus signs refer to instantons and anti-instantons,
respectively. The closure of the three-form field strengths then implies
Laplace-like equations for the scalars. On a flat spacetime ${\cal M}$ with
points $\{x_i\}$ excised from $\fieldR^4$, we find the multi-centered
solutions in terms of two harmonic functions,
 \begin{equation} \label{5-inst-sol}
  \e^{-\phi} = \e^{-\phi_\infty} + \sum_i\, \frac{|Q_{2i}|}{4\pi^2\,
  (x-x_i)^2}\ ,\qquad \e^{-\phi} \chi = \e^{-\phi_\infty} \chi_{\infty}
  + \sum_i\, \frac{Q_{1i}}{4\pi^2\, (x-x_i)^2}\ ,
 \end{equation}
where $Q_{1i}$, $Q_{2i}$, $\chi_\infty$, and $\phi_\infty$ are
independent integration constants; the latter two determine the
asymptotic values of the fields at infinity. We wrote the absolute value
of $Q_{2i}$ to make $\e^{-\phi}$ positive everywhere in space, and we
identify the string coupling constant via $g_s=\e^{-\phi_\infty/2}$.
Furthermore, two charges are defined by integrating the tensor field
strengths $H_{\mu\nu\rho\,I}=-\ep_{\mu\nu\rho\si}H^\si_I$ over 3-spheres
at infinity,
 \begin{equation}
  Q_I = \int_{S^3_\infty}\! H_I\ ,\qquad I=1,2\ .
 \end{equation}
They are related to the constants appearing in the scalar 
fields through the field equation (\ref{5-inst}). Using $**=-1$ on a 
three-form in four Euclidean dimensions, we find 
 \begin{equation}
  Q_2 = \mp \sum_i |Q_{2i}| \ ,\qquad Q_1 = \mp \sum_i Q_{1i}\ .
 \end{equation}
This implies that for instantons, $Q_2$ should be taken negative,
whereas for anti-instantons, $Q_2$ must be positive. Note that there is
no restriction on the sign of the $Q_{1i}$.

The (anti-) instanton action for the fivebrane can be computed using
the formulas in \cite{TV1}. It is finite only if $\chi$ remains finite
near the excised points $x_i$. In the limit, we find
 \begin{equation}
  \chi_i \equiv \lim_{x\rightarrow x_i} \chi(x) =
  \frac{Q_{1i}}{|Q_{2i}|}\ ,
 \end{equation}
which is finite whenever $Q_{2i}\neq 0$ for nonvanishing $Q_{1i}$.
This implies that the integrated Heisenberg invariants vanish,
$\hat{Q}_{1i}\equiv Q_{1i}-\chi_i |Q_{2i}|=0$. Plugging the solution into
the action, we find
\begin{equation}\label{inst-act}
 S_\mathrm{inst} =  \frac{|Q_2|}{g_s^2} + \half \sum_i \,|Q_{2i}| \,
 (\chi_\infty -\chi_i)^2\ .
\end{equation}
The quadratic dependence on the string coupling constant is precisely what 
corresponds to a wrapped NS-fivebrane \cite{BBS}.

For a single-centered instanton around $x_0$, this reduces to
 \begin{equation}\label{1inst-act}
  S_\mathrm{inst} = \frac{|Q_2|}{g_s^2}\, \Big(
  1 + \half \,g_s^2 (\Delta\chi)^2 \Big)\ ,
 \end{equation}
where $\Delta\chi=\chi_\infty-\chi_0$ is the difference between the
values of the R-R scalar at the boundaries $S^3_\infty\cup S^3_0$ of
$\fieldR^4-\{x_0\}$. It would be interesting to have a better string
theoretic interpretation of this term. 

An important issue is whether our solutions (\ref{5-inst-sol}) preserve
half of the supersymmetry. This question is analyzed in detail in
\cite{DTV}. It turns out that the spherically symmetric solutions are
BPS in the sense of preserving four supercharges. Interestingly, the
general multi-centered solution satisfying the Bogomol'nyi bound
(\ref{5-inst}) is not, and imposing the BPS condition leads to the
further restriction that all the $\chi_i$ must be equal. This implies
that the solution is characterized in terms of a single harmonic
function, given by $\e^{-\phi}$. When all $\chi_i$ are equal, the value
of the action is lowered and coincides with the spherically symmetric
case (\ref{1inst-act}). A possible interpretation is that the general
multi-centered solution is metastable and decays into the state where
all $\chi_i$ are equal. In such a state, the points $x_i$ can be brought
together to a spherically symmetric configuration without changing the
action. We call the general solution (\ref{5-inst-sol}) near-extremal
when the values of $\chi_\infty-\chi_i$ are all close to the lowest
value $\Delta\chi$.

\section{Membrane Instantons}

The second Bogomol'nyi equation that can be derived from
(\ref{DTM-action}) contains an arbitrary constant $\chi_0$ \cite{TV1},
 \begin{equation}\label{Bog-mem}
  \begin{pmatrix}H_{\mu1} \cr H_{\mu2}\end{pmatrix} = \pm \frac{1}{|\tau'|} 
\begin{pmatrix}
  \chi (\chi - \chi_0) \p_\mu \e^{-\phi} + \e^{-\phi} (\chi + \chi_0)
  \p_\mu \chi + 2 \e^{\phi} \p_\mu \e^{-\phi} \cr (\chi - \chi_0)
  \p_\mu \e^{-\phi} + 2 \e^{-\phi} \p_\mu \chi\end{pmatrix}\ ,
 \end{equation}
where $\tau'=(\chi-\chi_0)+2\I\e^{\phi/2}$. Below, for the
single-centered solution, we will identify $\chi_0$ with the value of
$\chi$ at the excised point. To solve these equations, we first consider
the combination
 \begin{equation}
  H_{\mu1} - \chi_0 H_{\mu2} = \pm \p_\mu (\e^{-\phi} |\tau'|)\ .
 \end{equation}
The Bianchi identities then imply that $h=\e^{-\phi} |\tau'|$ is
harmonic and positive everywhere. With $\chi$ expressed in
terms of $h$ and $\phi$, 
\begin{equation}\label{sol-chi}
\chi-\chi_0 = \e^{\phi}{\sqrt{h^2 - 4\e^{-\phi}}}\ ,
\end{equation}
the condition for $H_{\mu2}$ turns into
 \begin{equation}\label{H2}
  H_{\mu2} = \pm \frac{1}{\sqrt{h^2 - 4\e^{-\phi}\,}}\, \big( 2
  \e^{-\phi} \p_\mu h - h \p_\mu \e^{-\phi} \big)\ .
 \end{equation}
Notice that we have taken the positive branch in (\ref{sol-chi}), the
negative branch just changes the sign in (\ref{H2}). The Bianchi
identity and the harmonic property of $h$ now imply that
 \begin{equation}\label{f''}
  (h^2 - 4\e^{-\phi})\, \p_\mu\p^\mu \e^{-\phi} + 2 \p_\mu\e^{-\phi}\,
  \p^\mu \e^{-\phi} - 2h\, \p_\mu h\, \p^\mu e^{-\phi} + 2\e^{-\phi}\,
  \p_\mu h\, \p^\mu h = 0\ .
 \end{equation}

It is unclear how to solve this equation in general, but some
multi-centered solutions were constructed in \cite{M}. To continue, we
assume that the dilaton depends on the coordinates only through $h$.
This assumption is justified if we restrict ourselves to spherically
symmetric configurations,
 \begin{equation}\label{mem-harm}
  h = \e^{-\phi_\infty} |\tau'_\infty| + \frac{|\hat{Q}_1|}{4\pi^2\,
  (x-x_0)^2}\ ,
 \end{equation}
where $\hat{Q}_1=Q_1-\chi_0 Q_2$, which must be taken negative for
instantons (upper sign in (\ref{Bog-mem})). We now proceed to find the
most general spherically symmetric solution of (\ref{f''}), extending
the special cases of \cite{TV1}. By differentiating (\ref{f''}) once
more, one can solve for the dilaton in terms of three integration  
constants,
 \begin{equation}\label{dil}
  \e^{-\phi} = ah^2 +bh + c\ .
 \end{equation}
Combining this with equations (\ref{H2}) and (\ref{f''}), we find that
$c=-\beta^2$, where $\beta\equiv\pm Q_2/|\hat{Q}_1|$, and 
$b=-\beta\sqrt{1-\alpha}$, where $a=\alpha/4$. Positivity of the dilaton and
reality of $H_2$ requires furthermore $0\leq\alpha\leq 1$, but the case 
$\alpha=0$ must be treated separately and in fact precisely coincides
with the fivebrane instanton solution. 
Plugging (\ref{dil}) into (\ref{sol-chi}), and evaluating it at infinity,
one finds that $\alpha$ is fixed in terms of the charges and the 
asymptotic values of the fields,
\begin{equation}\label{alpha}
 \alpha = 1 - \frac{(\Delta\chi - 2\beta\, \e^{\phi_\infty})^2}
 {|\tau'_\infty|^2}\ ,
\end{equation}
where $\Delta\chi=\chi_\infty-\chi_0$. 
The solution for $\chi$ can be read off from (\ref{sol-chi}) and one can
check that $\chi_0$ is indeed the value at the origin. An important 
difference with the fivebrane is that $\chi$ needs a source at the excised 
point. 
Positivity of $\alpha$ forces, for fixed $g_s$ and  
$\Delta \chi$ (positive), $\beta$ to be in the
interval 
 \begin{equation}\label{window}
  \frac{\Delta\chi - |\tau'_{\infty}|}{2\,\e^{\phi_\infty}}\, \leq\,
  \beta\, \leq\, \frac{\Delta\chi + |\tau'_{\infty}|}
  {2\,\e^{\phi_\infty}}\ .
 \end{equation}

The complete solution is given in terms of one harmonic function, and
all the constants $a$, $b$, $c$ are fixed in terms of the charges and
the asymptotic values of the scalars. We have checked, using the
supersymmetry transformation rules given in \cite{TV2,DTV}, that the
solution preserves half of the supersymmetry (except in the limit 
$\alpha \rightarrow 0$), and 
hence is truly BPS. The instanton action can be computed using the formulas 
of \cite{TV1}, and equals
 \begin{equation}
  S_\mathrm{inst} = |\tau'_\infty| \,\Big( |\hat{Q}_1| + \frac{1}{2}\,
  \Delta\chi\, Q_2 \Big)\ .
 \end{equation}
Positivity of the action is guaranteed by the bounds (\ref{window}).
For $Q_2=0$, which trivially satisfies (\ref{window}), we 
find instanton actions which are, for $\Delta \chi=0$,
inversely proportional to the string coupling 
constant, $S_\mathrm{inst} = 2|{Q}_1|/g_s$.
This is the correct $g_s$-dependence of a membrane 
instanton with charge ${Q}_1$ \cite{BBS}. Adding fivebrane 
charge $Q_2$ raises the action until, for fixed $g_s$ and $\Delta \chi$, one 
reaches the $\alpha\rightarrow 0$ bound (\ref{window}). At that point, the 
solution is no longer stable, and should be replaced by the fivebrane 
instanton.

\medskip

\textbf{Acknowledgement} 

S.V.\ would like to thank the organizers of the Simons Workshop in
Mathematics and Physics, Stony Brook 2003, and those of the Ahrenshoop
Symposium, Wernsdorf 2003, for their hospitality and stimulating meetings. We
have also benefited from discussions with many of the participants.


\end{document}